\documentclass{WileyMSP-template}

\usepackage{cite}
\usepackage{xcolor}

\usepackage{color}
\newcommand{\blu}[1]{{\color{black}{#1}}}

\begin{document}

\pagestyle{fancy}
\rhead{\includegraphics[width=2.5cm]{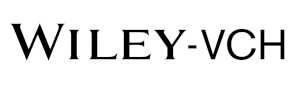}}

\title{Coalescence of Sessile Polymer Droplets: A Molecular Dynamics Study}

\maketitle


\author{Soheil Arbabi}
\author{Panagiotis E. Theodorakis*}


\begin{affiliations}
S. Arbabi, P. E. Theodorakis\\
Institute of Physics, Polish Academy of Sciences, Al. Lotnik\'ow 32/46, 02-668 Warsaw, Poland\\
Email Address: panos@ifpan.edu.pl
\end{affiliations}



\begin{abstract}
Droplet coalescence is ubiquitous
in nature and the same time key to various 
technologies, such as inkjet printing. 
Here, we report on the coalescence of polymer
droplets with different chain lengths coalescing
on substrates of different wettability.
By means of molecular dynamics simulations
of a coarse-grained model, \blu{it is found}
that the rate of bridge growth is higher
in the case of droplets with smaller contact angles
(more wettable substrates) and decreases with the
increase of the chain length of the polymers. 
Different behavior has also been
identified in the dynamics of the approach of 
the two droplets during coalescence with 
the substrate wettability playing a more important
role compared to the chain length of the polymers.
While the dynamics of the droplet are greatly affected
by the latter parameters, the density profile and
flow patterns remain the same for the different cases.
Thus, we anticipate that our work provides further
insights into the coalescence of liquid polymer 
droplets on solid substrates with implications 
for relevant technologies.
\end{abstract}


\section{Introduction}\label{intro}

Droplet coalescence is ubiquitous in nature and
at the same time much relevant for various technologies,
such as spraying and printing \cite{Eggers2008,Wijshoff2010},
where the rate of this process
can determine the efficiency of the application.
The primary factor controlling coalescence is the interplay
of viscous and inertial forces as droplets minimise their
total surface-tension energy by coalescing \cite{Verdier2002}. 
Despite research in this area \cite{Paulsen2014,yoon2007coalescence,khodabocus2018scaling,perumanath2019droplet,eggers1999coalescence,aarts2005hydrodynamics,Sprittles2012},
droplet coalescence remains a fascinating phenomenon with
many of its aspects calling for further \blu{investigations} 
to reach adequate understanding of this phenomenon in various scenarios \cite{Paulsen2014,yoon2007coalescence,khodabocus2018scaling,perumanath2019droplet,eggers1999coalescence,aarts2005hydrodynamics,Sprittles2012}.
On the one hand, part of this gap in knowledge is due to device limitations,
since experiments cannot capture the initial fast stages of
droplet coalescence. On the other hand,
the singularity at the contact point during the initial stages
of coalescence presents challenges for numerical modelling \cite{eggers1999coalescence,duchemin2003inviscid,Sprittles2012}, 
despite progress in this area \cite{Sprittles2012},
while hydrodynamic models are only applicable at the later
stages of 
coalescence \cite{yeo2003film,hu2000drop,mansouri2014numerical}.
Apart from device and methodology limitations,
further understanding at molecular scales is much desirable as 
applications often require greater control at nanoscales. 
Moreover, the role of a substrate
in the coalescence of sessile droplets deserves further
consideration
despite research in this area by theory and 
experiment \cite{Hernandez2012,Eddi2013,Lee2012,Ristenpart2006,Nikil2007}, 
especially in the context of complex liquids containing
various additives, such as polymers and colloids.

Droplet coalescence
takes place in three stages, \blu{with the first being}
the initial droplet approach when the droplets are close
enough to interact with each other and form in between
the so-called bridge. 
Then, the bridge-growth stage follows, which eventually
results in the final reshaping of the two droplets towards
the equilibrium state of a single spherical-cap droplet, 
which is the state of minimum energy. 
From the perspective of fluid dynamics the initial 
bridge growth is generally driven by viscous forces,
as a result of the interactions between molecules,
while inertial forces dominate the
coalescence process at the later 
stage \cite{duchemin2003inviscid,eggers1999coalescence}.
In the case of the viscous regime, a linear scaling in time 
$b \propto t$ has been suggested for the bridge radius, $b$, 
or logarithmic corrections $t\ln t$, 
while a scaling $b\propto \sqrt{t}$
has been proposed for the inertial 
regime \cite{eggers1999coalescence}.
However, the dynamics of the bridge growth
is still under debate, for example, 
an inertially limited viscous regime has been 
reported \cite{paulsen2012inexorable,paulsen2013approach}
or the proposition of a modified Ohnesorge number in the case of
immiscible droplets for coupling the linear 
and power-law scalings \cite{xu2022bridge}. 
All-atom molecular dynamics (MD)
\blu{simulations} \cite{perumanath2019droplet} 
have described the initial stage of the bridge growth 
\blu{for water droplets},
not achievable by continuum simulation or experiment. 
In particular, the formation 
of multiple precursor bridges at the pinch point were observed,
which result from thermal capillary waves that exist
at the droplets' surface. In this case, simulations
suggest that multiple bridges that expand linearly
in time develop at the surface and the transition 
to the classical hydrodynamics regime only takes 
place when the bridge radius exceeds a thermal length
defined as 
$l_T  \approx {\left ( k_BT/\gamma \right) }^{1/4} R^{/1/2}$,
where $k_B$ is Boltzmann's constant, $T$ the temperature, 
$\gamma$ the liquid--vapour (LV) surface tension, and
$R$ the radius of the droplets.

In the case of droplet coalescence on solid substrates,
much less is known, despite the immediate implications
for applications, for example, in 
coatings \cite{ristenpart2006coalescence} 
and microfluidics \cite{feng2015advances} technologies.
In particular, in the case of coalescence
of low-viscosity droplets on a substrate
it has been \blu{experimentally}
found that the the bridge height, $b$,
grows with time as $t^{2/3}$ when the contact angle is below
90$^\circ$, while a scaling $b \propto t^{1/2}$ has
been observed for contact angles above
90$^\circ$ \cite{Eddi2013},
which is the scaling found in the inertial regime 
for freely suspended drops \cite{eggers1999coalescence,ristenpart2006coalescence}.
Moreover, a geometrical model that unifies the inertial 
coalescence of sessile and freely suspended drops and
can capture the transition from the $2/3$ to the $1/2$ exponent 
in the case of sessile droplets has been proposed \cite{Eddi2013}.
In addition, in the case of asymmetric coalescence, that is
droplets with different contact angles, the theory predicts
that the shape of the bridge can be described by similarity solutions
of the one-dimensional lubrication theory, with the bridge 
growing linearly in time and 
exhibiting dependence on the contact
angles \cite{Hernandez2012}.
In earlier experimental studies, 
a power-law growth at early times as $t^{1/2}$
has been suggested for the symmetric case, while the growth rate
appeared to be sensitive to both the radius and the height of the
droplet with a scaling $H/R$, where $H$ is the height of the droplet
from the substrate to its apex and $R$ its 
radius \cite{ristenpart2006coalescence}. 
Further experimental work on droplets with contact angles in
the range 10$^\circ$--56$^\circ$
has found that the bridge growth scales as a power
law with exponents in the range $0.5061$ to $0.8612$ with
data deviating from the power law at longer times during coalescence
for contact angles larger than 24$^\circ$. Moreover, 
a power law with an exponent $0.2901$ has been found
for the width of the bridge \cite{Lee2012}. 
Finally, further experimental work has focused on analysing the
morphology and dynamics of droplet coalescence on
substrates \cite{Kapur2007}.

Despite previous work on the coalescence of sessile droplets, 
many aspects of this phenomenon require further investigation.
One of them is the role of viscosity in the coalescence 
for substrates with different wettability.
Viscosity is expected to play
a role, especially in the context of polymer
droplets studied here, where in addition to surface-tension-effects
differences, entanglement effects may also play
a role for longer polymer chains or the 
polymer--polymer interactions close to the contact
line in both the initial and later stages of
coalescence. Here, we attempt to elucidate these 
points and fill in the gap in knoweledge in this area
by carrying out molecular dynamics simulations of
a coarse-grained model for droplets comprised
of polymer chains with different length 
on substrates of different wettability with
equilibrium contact angles of individual droplets 
above and below 90$^\circ$.
We find that the bridge length dynamics are much
slower in the case of polymer droplets than what
is observed for water droplets. Moreover, the 
coalescence process considerably slows down with the
increase of polymers chain \blu{length}. Furthermore, 
more wettable substrates have consistently faster 
bridge growth dynamics in comparison with the less
wettable substrates. The wettability of the substrate also
affects significantly the dynamics 
of the bridge angle and the approach of
the coalescing droplets, while the viscosity of the droplets
appears to have a smaller effect.
In the following, we describe our simulation model and method. 
Then, we discuss our results, while in the final section we
draw our conclusions.

\section{Simulation model and \blu{methods}}

\begin{figure}
\centering
\includegraphics[width=0.7\textwidth]{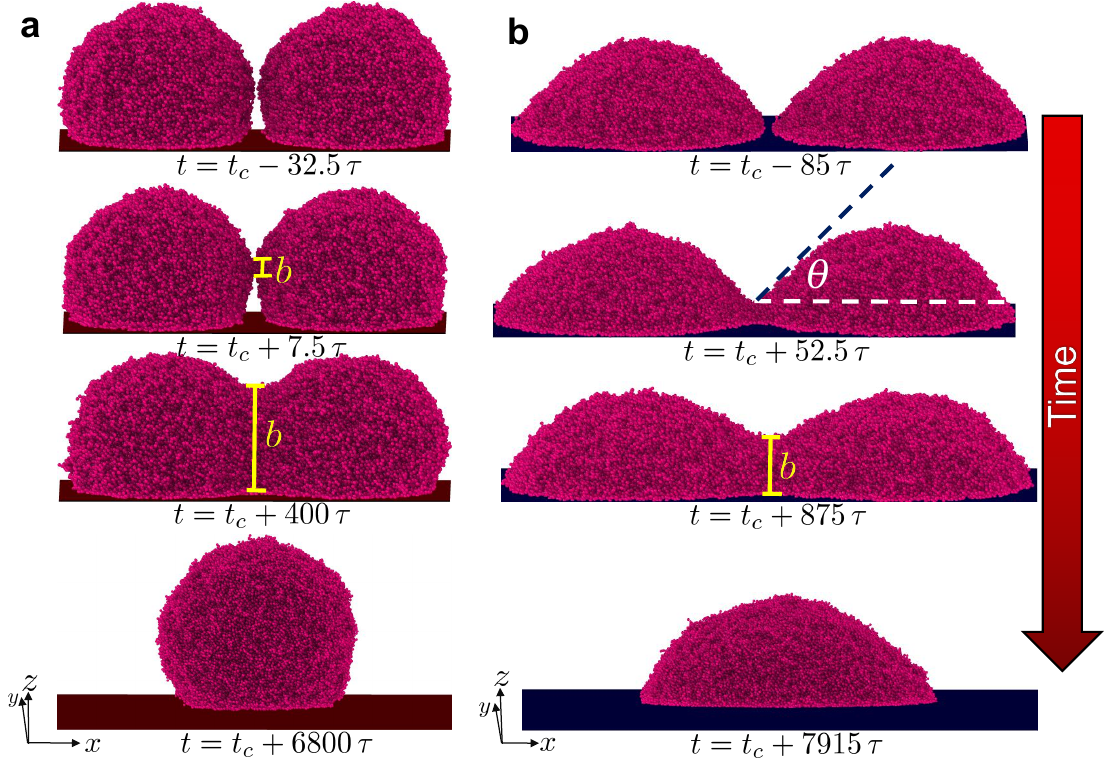}
\caption{\label{fig:1} Evolution of the droplet
coalescence on a solid substrate 
with a lower
(a, $\varepsilon_{\rm pw} = 1.1~\epsilon$)
and a higher (b, $\varepsilon_{\rm pw} = 2.5~\epsilon$) 
wettability as a function of time, \blu{$t$, from the
first permanent contact of the droplets at time $t_c$}, as indicated. 
Here, $N=10$ beads. \blu{Moreover, the angle, $\theta$, and}
the bridge length, $b$, are indicated.
The implicit, smooth substrate modelled by the 9--3 LJ
potential of Equation~\ref{eq:LJpotential93} 
is illustrated by a solid colour.}
\end{figure}

Our system consists of two polymer droplets
placed next to each other as shown in \textbf{Figure~\ref{fig:1} }
to initiate their coalescence. Each droplet contains
 polymer chains with 
the same number of monomers (beads), \blu{$N$}. 
The polymer chains are modelled by
the standanrd bead--spring
model \cite{Murat1989,Theodorakis2017,Theodorakis2022},
where all beads interact with a truncated
and shifted Lennard-Jones (LJ) potential,
\begin{equation}\label{eq:LJpotential}
U_{\rm LJ}(r) = 4\varepsilon_{\rm ij} \left[  \left(\frac{\sigma_{\rm  ij}}{r}
\right)^{12} - \left(\frac{\sigma_{\rm ij}}{r}  \right)^{6}    \right].
\end{equation}
This interaction is applied for beads within the cutoff
distance $r_{\rm c}=2.5~\sigma$, where $\sigma$ is
the length unit. The interaction between polymer beads \blu{is}
$\varepsilon_{\rm pp}=\epsilon$, where $\epsilon$
is the unit of energy. The temperature 
\blu{of the system is}, $T=\epsilon/k_B$,
where $k_B$ is Boltzmann's constant. 
Moreover, consecutive beads along a polymer chain
are tethered by the ``finitely extensible nonlinear elastic''
(FENE) potential,
\begin{equation}\label{eq:KG}
 U_{\rm FENE}(r) = -0.5 K_{\rm FENE} R_{\rm 0}^2 \ln \left[ 1 - \left(\frac{r}{R_{\rm 0}} \right)^2  \right],
\end{equation}
where $r$ is the distance between two 
consecutive beads along the polymer chain,
$R_{\rm 0}=1.5~\sigma$ expresses 
the maximum extension of the bond, and
$K_{\rm FENE} = 30~\epsilon/\sigma^2$
is an elastic constant. 
The length of the polymer chain in this model in effect
varies the viscosity of the droplets \cite{Kajouri2023}.
Here, the chain length is the same for both droplets
in each system and is chosen in the range $N=10-640$ beads. 
\blu{Since the total number of beads in each droplet is 57600, 
using longer chains would also require the increase of the overall
size of the droplet, in order to ensure that the majority of the
chains are not on the surface of the droplet, thus avoiding artifacts
that may not apply in macroscopic droplets. Moreover, increasing
$N$ and the total number of beads in the droplets
would result in longer times required for the equilibration of the
initial droplets and the coalescence experiments to reach the final
equilibrium stage. Still, it would be valuable to extend the range
of $N$ in future investigations and carry out a full scaling analysis 
of droplet properties on the chain length $N$ and the overall droplet size.}
The wettability of the substrate by the droplet is controlled
through the parameter $\varepsilon_{\rm pw}$
of the 9--3 LJ potential, which 
describes the interaction
of the polymer beads with an implicit, smooth wall \cite{Theodorakis2021materials},
\begin{equation}\label{eq:LJpotential93}
U_{\rm w}(z) = 4\varepsilon_{\rm pw} \left[  \left(\frac{\sigma_{\rm  s}}{z}
\right)^{9} - \left(\frac{\sigma_{\rm s}}{z}  \right)^{3}    \right],
\end{equation}
where $z$ is the normal (vertical) distance of the beads from
the substrate within a cutoff distance $z_{\rm c}=2.5~\sigma$. Here, $\sigma_{\rm s}=\sigma$.

To evolve our system in time and control the temperature
of the system, the Langevin thermostat is used as done 
previously \cite{Theodorakis2011}. 
The equation of motion for the coordinates $\{r_i(t)\}$ of 
the beads of mass $m$ ($m$ is the unit of mass)
\begin{equation}\label{Eq3}
 m\frac{d^{2}\textbf{r}_{i}}{dt^{2}}=-\nabla U_{i}- 
\gamma \frac{d\textbf{r}_{i}}{dt}+\Gamma_{i} (t).
\end{equation}
is numerically integrated for each bead 
using the LAMMPS package \cite{Plimpton1995}. In
Equation~\ref{Eq3}, $t$ denotes the time, $U_{i}$ is the total potential acting on the $i$th
bead, $\gamma$ is the friction coefficient, and ${\Gamma}_i(t)$ is the random force. As
is well-known, $\gamma$ and $\Gamma$ are related by the usual fluctuation--dissipation relation
\begin{equation}\label{Eq4}
<\Gamma_{i}(t).\Gamma_{j}(t^{'})>=6k_{B}T\gamma\delta_{ij}\delta(t-t^{'}).
\end{equation}
Following previous work\cite{Grest1993,Theodorakis2011,Murat1991}, 
the friction coefficient was
chosen as $\gamma=0.5~\tau^{-1}$. Equation~\ref{Eq3} 
was integrated using an integration time step
of $\Delta t=0.01~\tau$, where
the time unit is $\tau=(m\sigma^{2}/\epsilon)^{1/2}$. 
A single droplet is first
equilibrated for adequate time, so that the 
total energy has reached a minimum and properties,
such as the mean contact angle and average shape of the droplet
do not change with time. Then, the equilibrated droplet
is cloned and positioned on the substrate as shown in
Figure~\ref{fig:1}. In this case, the size
of the box is chosen such to accommodate the
two droplets avoiding the interaction of 
mirror images of the droplets due to the 
presence of periodic boundary conditions in
all Cartesian directions. Moreover, the
use of polymer droplets leads to the
absence of vapour in the system \cite{Tretyakov2013},
which greatly facilitates \blu{the
analysis of the trajectories and maintaining
the same thermodynamic conditions during the
simulation of either the individual droplet or the two 
coalescing droplets.}
Different scenarios of substrate
wettability were considered in our study, for which 
$\varepsilon_{\rm pw}$ is
$2.5~\epsilon$ or $1.1~\epsilon$ .
In this case, the equilibrium contact angles of the
individual droplets before coalescence are 78$^\circ$
and 118$^\circ$, respectively. \blu{To
estimate the contact angle of the droplet,
a method that avoids a fitting procedure is used,
which has been described in 
Reference~\cite{theodorakis2015modelling}}. 
We have also found that the equilibrium contact angles
of the individual droplets do not show any statistically
significant dependence on the length, $N$, of the polymer
chains.

\blu{To analyse the bridge growth dynamics, snapshots of 
the system are frequently dumped, especially for the
initial stage, typically every 250 integration time steps
and beads are assigned to a three-dimensional grid with
size $2.5\sigma$ in all directions. For the analysis of each snapshot,
the center of the bridge is located in the middle of
the grid, corresponding to the position $x=0$ in the
$x$ direction of the coordinate system
and any rotation of the droplets around
the $z$ axis has been removed. 
This facilitates our analysis and guarantees
that our measurements of the bridge radius
and all other properties (\textit{e.g.} density profiles)
remain consistent as
coalescence proceeds. The snapshots 
of Figure~\ref{fig:1} have been taken after
performing the above procedure, which is 
manifested by the perfect
alignment of the droplets along their long axis in the $x$ direction
and the bridge is also placed in the middle of the substrate
on the $x-y$ plane. 
The three dimensional grid is also used
to calculate the profile of the number density of the
droplets by considering a slab along
the $x-z$ plane in the $x$ direction that passes through
the center of the bridge. Further details regarding the
calculation of the various properties are provided
later during the discussion of the respective results.
}

\begin{figure}
\centering
\includegraphics[width=\textwidth]{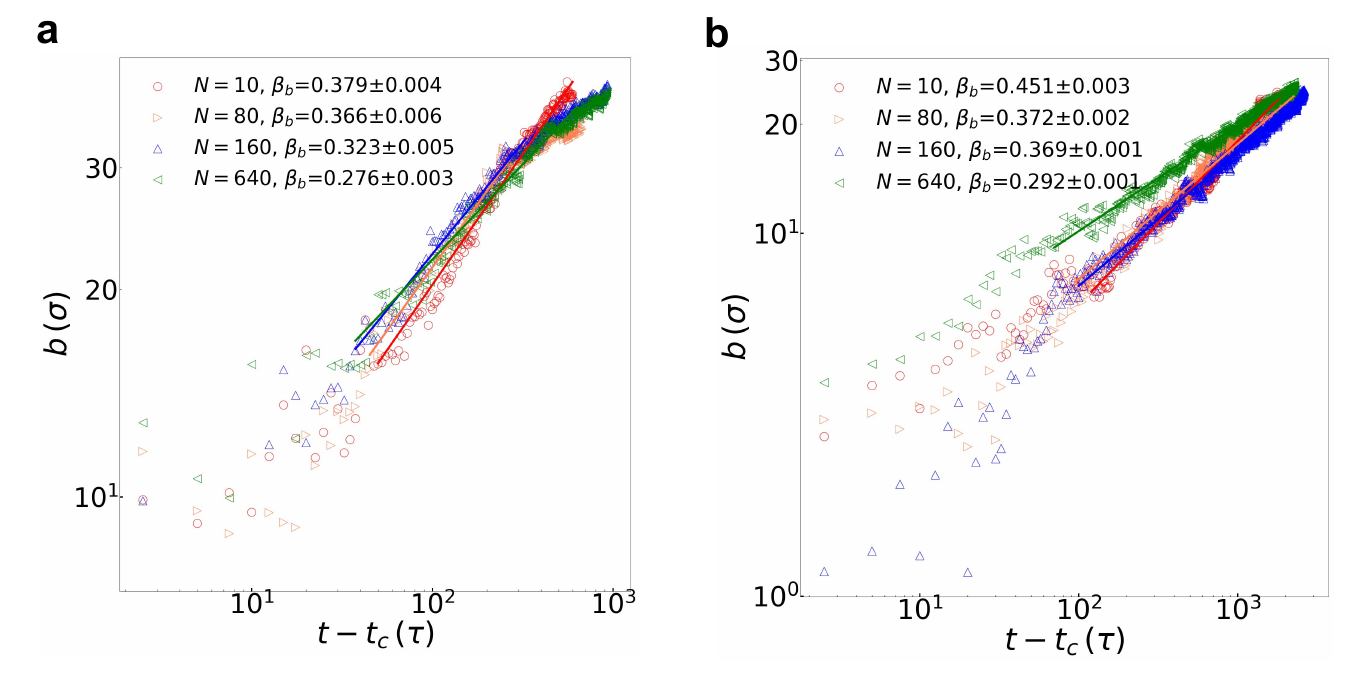}
\caption{\label{fig:2} Bridge length, $b$, as a function
of time, $t$, counting from the time of first permanent
contact of the coalescing droplets, $t_{c}$. Data for
polymer droplets with different chain length are shown, $N$,
as indicated.
(a) $\varepsilon_{\rm pw}=1.1~\epsilon$;
(b) $\varepsilon_{\rm pw}=2.5~\epsilon$.
The power-law exponents ($\sim t^{\beta_b}$) are reported
for the bridge length \blu{with values of  $\chi^2/\rm ndf \approx 1$, where ndf
indicates the number of degrees of freedom.}
}
\end{figure}

\section{Results and Discussion}
Figure~\ref{fig:1} shows typical coalescence cases on 
substrates with different wettability, corresponding
to contact angles of lower and greater than 90$^\circ$.
A key parameter for characterising the dynamics
of the coalescence process is the bridge length, $b$,
which is indicated for each case in Figure~\ref{fig:1}.
When the substrate is less wettable (contact angle
greater than 90$^\circ$), the bridge initially forms
above the substrate at the contact point
between the LV interface of the coalescing droplets,
and later comes into
contact with the substrate as the
coalescence process proceeds (Figure~\ref{fig:1}a).
In contrast, 
in the case of more wettable substrates (
contact angles lower than 90$^\circ$),
the bridge grows
onto the substrate from the beginning of the 
coalescence process. 
While the time that bridge is in contact with the
substrate is expected to affect the dynamics of the
droplets, in the case of more wettable substrates
the interaction between the droplet and the substrate
is also stronger.

\begin{figure}
\centering
\includegraphics[width=0.5\textwidth]{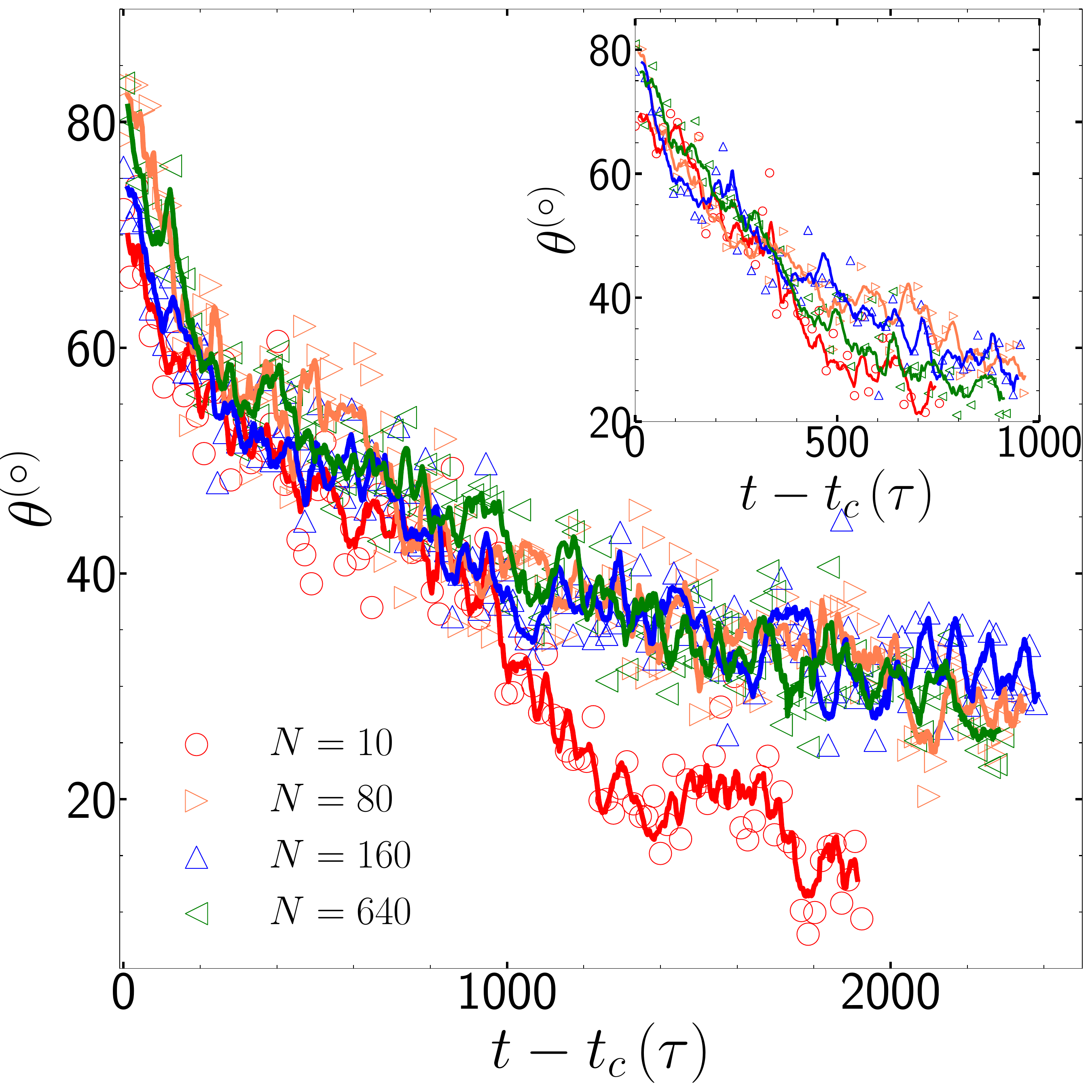}
\caption{\label{fig:3} Angle $\theta$ (see Figure~\ref{fig:1}
\blu{and main text for further details})
as a function of time, $t$, counting from the time of first 
permanent contact of the coalescing droplets, $t_{c}$.
Data for polymer droplets with different chain length, $N$,
are shown, as indicated. The lines are a guide for the eye.
Here, $\varepsilon_{\rm pw}=2.5~\epsilon$ 
and  $\varepsilon_{\rm pw}=1.1~\epsilon$ (inset). 
}
\end{figure}

\textbf{Figure~\ref{fig:2}} presents our results 
for the dynamics of the
bridge length, $b$, on the two different substrates. 
Apart from the initial thermal 
regime \cite{perumanath2019droplet},
we find that in terms of the
bridge length the dynamics of coalescence on both substrates
can be described by a power-law behaviour ($\sim t^\beta$) 
with exponents that are clearly lower than 1/2 
(contact angles greater than 90$^\circ$) and 2/3
(contact angles lower than 90$^\circ$), which have been
reported for water droplets \cite{Eddi2013}.
Therefore, our results suggest that
the rate of coalescence is slower in the case of 
polymer droplets compared to the case of water droplets.
Moreover, the increase of the polymer
chain length leads to gradually 
decreasing values of the power-law
exponent for both types of substrates.
However, exponents are clearly higher in the case of 
the more wettable substrate, which suggests that the 
coalescence process be faster when the attaction
of the polymer chains to the substrate is stronger. 
Hence, an increased substrate wettability
appears to accelerate the dynamics of the bridge growth,
thus facilitating droplet coalescence throughout the
range of $N$ studied here. Moreover, we have identified
the presence of a second regime at the final stages of the
coalescence process and when almost the bridge has been
fully developed in the case of less
wettable substrates, an effect that is more pronounced
for longer chain lengths $N$. 
In summary, we find that an increasing chain length of
the droplets will slow down the coalescence of polymer 
droplets and more wettable substrates will exhibit faster
dynamics than less wettable substrates with power-law
exponents, $\beta_{b}$, significantly lower than 
what has been observed for sessile water droplets.

\textbf{Figure~\ref{fig:3}} presents results for the angle
$\theta$ at the bridge (see Figure~\ref{fig:1}).
A symmetric angle is defined for the second
droplet \blu{of Figure~\ref{fig:1}}  
and the average of the two angles for each
snapshot is considered as the value of the
angle $\theta$. To calculate the angle $\theta$, one
considers a horizontal plane that passes through
the top of the bridge. Then, the angle is calculated
based on the curvature of the droplets as discussed
in a previous study, 
thus avoiding a fitting 
procedure \cite{theodorakis2015modelling}.
\blu{Estimating the angles can in general
be highly sensitive to the details
of the definition of a sharp interface, 
as well as the fitting procedure \cite{yatsyshin_kalliadasis_2021,Mugele2002}.
Moreover, models that could account for the 
disjoining pressure effects,
for example, in the context of droplets on solid substrates,
might perform better than fitting spherical
caps to nanodroplets \cite{yatsyshin_kalliadasis_2021}.
}
In general, our data for the
angle $\theta$ appear
noisier than the data referring to the bridge length.
One of the main reasons for this are the
larger fluctuations on the droplets
shape during the coalescence process. 
Hence, a discussion here can only focus on
the dynamics of the angle $\theta$, which seems to
greater be affected at the earlier times
of coalescence in the case of more wettable
substrates, while curves seem to saturate
for chain lengths $N \geq 80$ beads. 
Moreover, a faster rate of change appears
in the case of the less wettable substrates. 

\begin{figure}
\centering
\includegraphics[width=\textwidth]{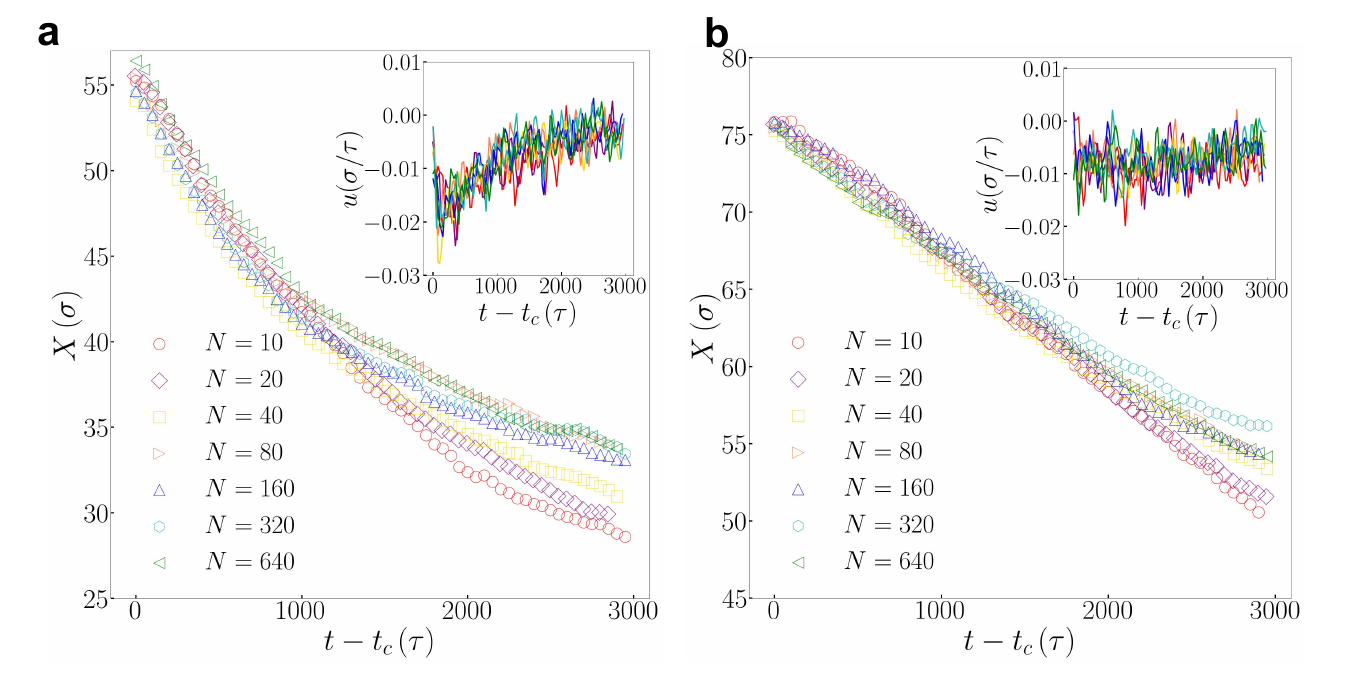}
\caption{\label{fig:4} Distance $X$ in the $x$ direction
between the centre of 
mass of coalescing droplets for
cases of different chain length, $N$, as indicated.
Insets show the instantaneous velocity of approach
$u=dX/dt$.
(a) $\varepsilon_{\rm pw}=1.1~\epsilon$;
(b) $\varepsilon_{\rm pw}=2.5~\epsilon$.
}
\end{figure}

\begin{figure}
\centering
\includegraphics[width=0.75\textwidth]{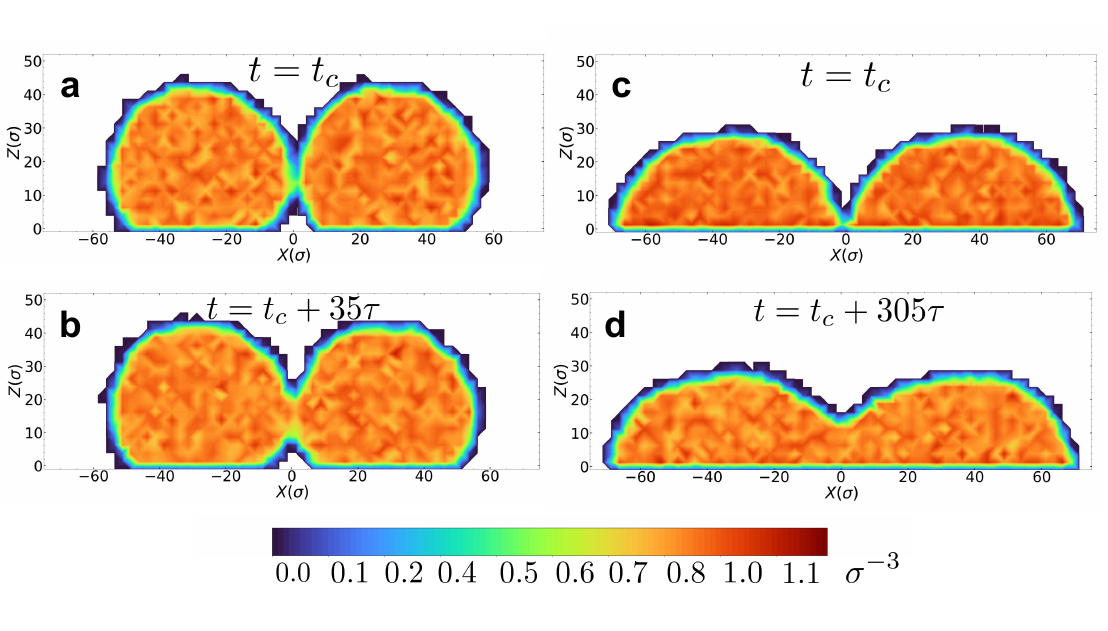}
\caption{\label{fig:5} Profiles of the number density along
a cross-section in the $x$ direction ($x-z$ plane)
of the coalescing droplets ($N=640$ beads)
at different stages (upper panels correspond
to snapshots obtained at time $t_c$, when the droplets
come into permanent contact).
(a, b) $\varepsilon_{\rm pw}=1.1~\epsilon$;
(c, d) $\varepsilon_{\rm pw}=2.5~\epsilon$.
}
\end{figure}

We have further explored the dynamics of the coalescence process
by monitoring the distance $X$ of the centre of mass of the
coalescing droplets, and, also, calculated its derivative with
time, which reflects the instantaneous velocity of approach of
the droplets (\textbf{Figure~\ref{fig:4})}. 
Our data for $\varepsilon_{\rm pw}=1.1~\epsilon$
(less wettable substrate) show two different dynamics regimes
with a transition between them that 
is more pronounced in the case of droplets with
longer polymer chains ($N \geq 40$ beads). 
This transition seems to become smoother as the chain length 
decreases. Moreover, the instantaneous velocity, $u$, of
the approach of the droplets is higher at the initial
stages of coalescence and then rather reaches a smaller
value, which remains constant until the bridge has fully
developed.
This velocity appears to be similar for the different
systems, independently of the chain length. 
In the case of the systems with $\varepsilon_{\rm pw}=2.5~\epsilon$,
a different behaviour is observed. 
$X$ steadily decreases, while the velocity, $u$, 
obtains small values over the entire coalescing process
with the initial instantaneous velocity
of the approach of the droplets to exhibit a slightly
higher (more negative velocity, since the distance $X$ decreases)
velocity. Hence, although the bridge growth dynamics is faster
in the case of the more wettable substrate, the approach
of the two droplets appears slower in the case of the
more wettable substrate.

Finally, we have calculated the density profiles of the
droplets during coalescence. From the obtained results,
we have not identified any noticeable changes in
the density profiles for droplets of different chain length
and substrates with different wettability.
We have also analysed the flow patterns and they have also
not revealed any noticeable differences for the
different systems. Typical density profiles for various cases
are presented in \textbf{Figure~\ref{fig:5}} at an initial stage
of the coalescence process, when droplets come into contact,
and at a later stage when the bridge has been clearly developed. 
Hence, while the dynamics of the coalescence process
depends on the choice of the substrate and the chain
length of the polymers, no noticeable changes in the patterns of 
the density and the flow are observed during coalescence
for the various cases considered in our study.

\section{Conclusions}
In this study, we have characterised the dynamics of
the coalescence of polymer droplets with different
chain lengths on substrates with
different wettability, where the contact angle of 
individual droplets is less and above 90$^\circ$. 
The rate of coalescence is a key property and 
can be characterised by the growth rate of the 
bridge length. We find that polymer droplets
overall show a slower rate of the bridge growth
in comparison with what has been observed in the case
of water droplets in experiments.
Moreover, the dynamics are slower as the length of the polymer chains
of the droplets increases. Also, we find
that more wettable substrates will exhibit faster
dynamics, which suggests that a stronger attraction
between the droplet and the substrate will accelerate
the bridge growth. In addition, we have characterised the dynamics
of the approach of the two droplets based on the
distance between the centre of masses of the coalescing
droplets. The behaviour is different when the wettability of
the substrate changes with two different regimes being
more pronounced in the case of less wettable substrates.
In this case, differences in the dynamics
between droplets with different chain lengths have
been also observed. 
While the dynamics of the coalescence can vary when
the length of the polymer chains or the substrate
wettability vary, the density and velocity profile
patterns do not reveal any dependence on these parameters.
Thus, we anticipate that our study provides insights
in the coalescence of liquid polymer droplets on solid substrates.









\medskip
\textbf{Acknowledgements} \par 
This research has been supported by the National Science Centre, Poland, 
under grant\\
No.\ 2019/34/E/ST3/00232. 
We gratefully acknowledge Polish high-performance 
computing infrastructure PLGrid (HPC Centers: ACK Cyfronet AGH) 
for providing computer facilities and support 
within computational grant no. PLG/2022/015261.

\medskip

%
\bibliographystyle{MSP}
\bibliography{biblio}

\newpage

\begin{figure}
\textbf{Table of Contents}\\
\medskip
  \includegraphics{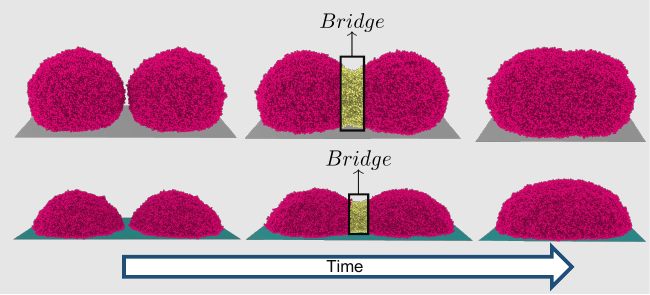}
  \medskip
  \caption*{Droplet coalescence of polymer droplets exhibits faster bridge growth for
  more wettable substrates (contact angle lower than $90^\circ$) and slower for droplets
  with longer polymer chains. The dynamics of droplets' approach
  are different for less wettable substrates (contact angle higher
  than $90^\circ$) with two different
  regimes identified as the growing bridge comes into contact with the substrate.} 
\end{figure}

\end{document}